\documentclass{mem}
\usepackage{natbib}\usepackage{txfonts}\usepackage{balance}
\usepackage{graphicx}
\usepackage[a4paper]{hyperref}
\idline{xxx}{1}
\begin{document}
\newcommand{\etal}{et\,al.\ }
\newcommand{\logg}{\mbox{$\log g$}}
\newcommand{\Teff}{\mbox{$T_\mathrm{eff}$}}
\newcommand{\lppr}{\stackrel{<}{\scriptstyle \sim}}
\newcommand{\lappr}{\raisebox{-0.4ex}{$\lppr $}}
\newcommand{\gppr}{\stackrel{>}{\scriptstyle \sim}}
\newcommand{\gappr}{\raisebox{-0.4ex}{$\gppr $}}
\newcommand{\msun}{\ensuremath{\, {\rm M}_\odot}}
\newcommand{\lsun}{\ensuremath{\,{\rm L}_\odot}} 

\title{
AGB star intershell abundances inferred from analyses of extremely hot
H-deficient post-AGB stars
}

   \subtitle{}

\author{
K. Werner\inst{1}, \, D. Jahn\inst{1}, \, T. Rauch\inst{1}, \, E. Reiff\inst{1}, 
\, F. Herwig\inst{2},
\, \and
\, J.W. Kruk\inst{3}
}

  \offprints{K. Werner}

\institute{
Institut f\"ur Astronomie und Astrophysik, Universit\"at T\"ubingen, Sand~1,
72076 T\"ubingen, Germany, \email{werner@astro.uni-tuebingen.de}
\and
Los
Alamos National Laboratory, Theoretical Astrophysics Group T-6, MS B227, Los
Alamos, NM 87545, U.S.A., \email{fherwig@lanl.gov}
\and
Department of Physics and Astronomy, Johns Hopkins University,
Baltimore, MD 21218, U.S.A., \email{kruk@pha.jhu.edu}
}

\authorrunning{Werner \etal}

\titlerunning{AGB star intershell abundances inferred from PG1159 stars}

\abstract{The hydrogen-deficiency in extremely hot post-AGB stars of spectral
class PG1159 is probably caused by  a (very) late helium-shell flash or a AGB
final thermal pulse that consumes the hydrogen envelope, exposing the usually-hidden
intershell region. Thus, the photospheric element abundances of
these stars allow to draw conclusions about details of nuclear burning and
mixing processes in the precursor AGB stars. We compare predicted element
abundances to those determined by quantitative spectral analyses performed with
advanced non-LTE model atmospheres. A good qualitative and quantitative
agreement is found for many species (He, C, N, O, Ne, F, Si) but discrepancies
for others (P, S, Fe) point at shortcomings in stellar evolution models for AGB
stars.
   } \maketitle{}

\begin{figure*}
\begin{center}
\includegraphics[scale=.50]{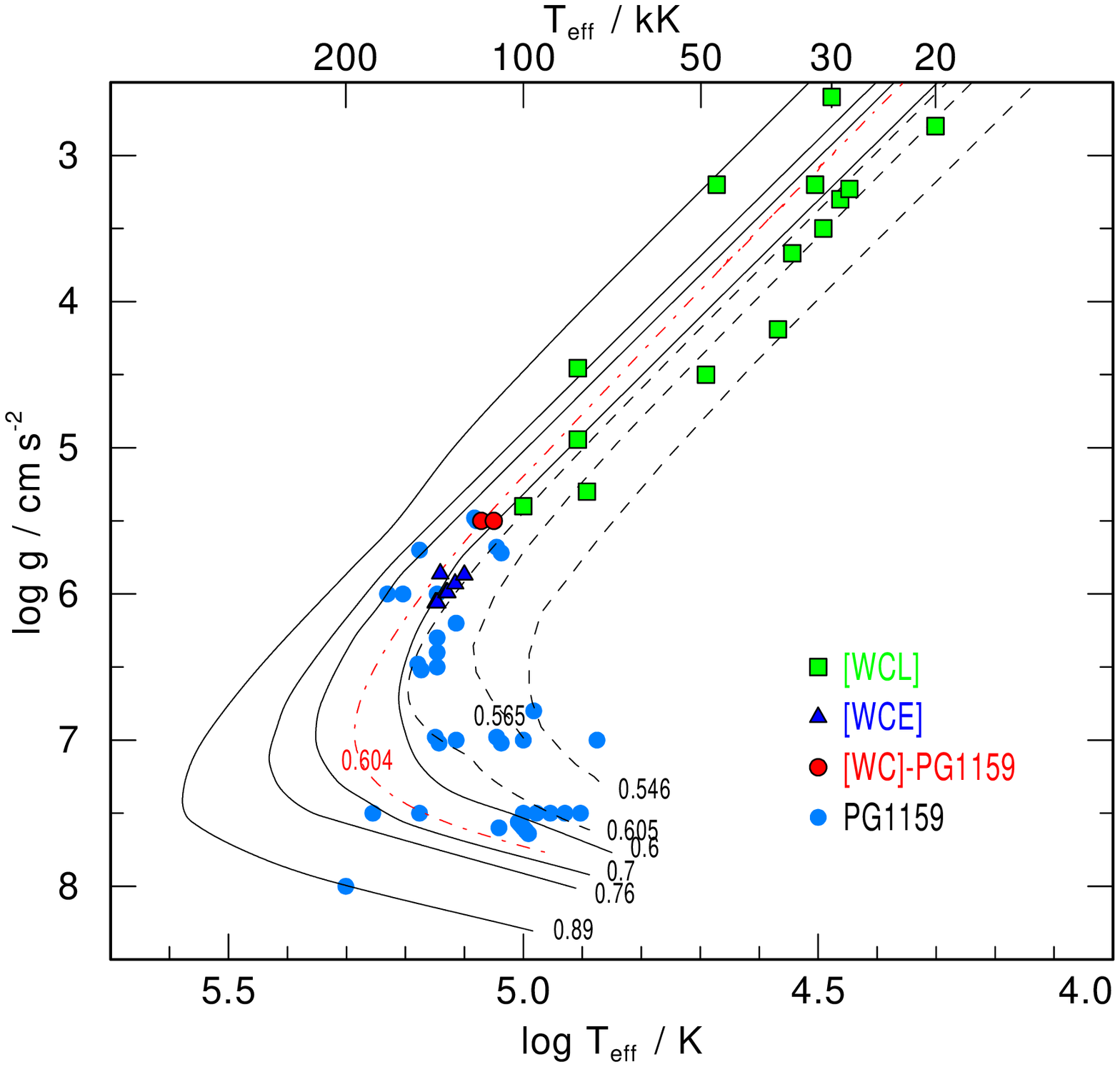}
\end{center}
\vspace{-0.5cm}
\caption{Hot hydrogen-deficient post-AGB stars in the
$g$--\Teff--plane. We identify Wolf-Rayet central stars of early and
late type \citep[{[WCE], [WCL],} from][]{hamann:97}, PG1159 stars
\citep[from][]{werner:06} as well as two [WC]--PG1159 transition objects
(Abell~30 and 78). Evolutionary tracks are from
\citet{schoenberner:83} and \citet{bloecker:95b} (dashed lines),
\citet{wood:86} and \citet{2003IAUS..209..111H} (dot-dashed line)
(labels: mass in M$_\odot$).  The latter $0.604\msun$ track is the
final CSPN track following a VLTP evolution and therefore has a
H-deficient composition. However, the difference between the tracks is
mainly due to the different AGB progenitor evolution.}\label{fighrd}
\end{figure*}
 
\section{Introduction}

The PG1159 stars are a group of 40 extremely hot hydrogen-deficient post-AGB
stars\footnote{For a recent review with a detailed bibliography see \citet{werner:06}.}. 
Their effective temperatures (\Teff) range between 75\,000~K and 200\,000~K. Many
of them are still heating along the constant-luminosity part of their post-AGB evolutionary
path in the HRD ($L \approx 10^4$\lsun) but most of them are already fading along
the hot end of the white dwarf cooling sequence (with $L$ $\gappr$
10\,\lsun). Luminosities and masses are inferred from spectroscopically
determined \Teff\ and surface gravity (\logg) by comparison with
theoretical evolutionary tracks. The position of analysed PG1159 stars in the
``observational HRD'', i.e. the \Teff--\logg\ diagram, are displayed in
Fig.\,\ref{fighrd}. The high-luminosity stars have low \logg\ ($\approx$\,5.5) while the
low-luminosity stars have a high surface gravity ($\approx$\,7.5) that is
typical for white dwarf (WD) stars. The derived masses of PG1159 stars have a mean of
0.62\,\msun, a value that is practically identical to the mean mass of WDs. The
PG1159 stars co-exist with hot central stars of planetary nebulae and the
hottest hydrogen-rich (DA) white dwarfs in the same region of the HRD. About
every other PG1159 star is surrounded by an old, extended planetary nebula.

What is the characteristic feature that discerns PG1159 stars from
``usual'' hot central stars and hot WDs? Spectroscopically, it is
the lack of hydrogen Balmer lines, pointing at a H-deficient surface
chemistry. The proof of H-deficiency, however, is not easy: The stars are very
hot, H is strongly ionized and the lack of Balmer lines could
simply be an ionisation effect. In addition, every Balmer line is blended by a
Pickering line of ionized helium. Hence, only detailed modeling of the spectra
can give reliable results on the photospheric composition. The high effective
temperatures require non-LTE modeling of the atmospheres. Such models for
H-deficient compositions have only become available in the early 1990s after new
numerical techniques have been developed and computers became capable enough.

The first quantitative spectral analyses of optical spectra from PG1159 stars
indeed confirmed their H-deficient nature \citep{werner:91}. It could be shown that
the main atmospheric constituents are C, He, and O. The typical abundance pattern
is C=0.50, He=0.35, O=0.15 (mass fractions). It was speculated that these stars
exhibit intershell matter on their surface, however, the C and O abundances were
much higher than predicted from stellar evolution models. It was further speculated
that the H-deficiency is caused by a late He-shell flash, suffered by the star
during post-AGB evolution, laying bare the intershell layers. The re-ignition of
He-shell burning brings the star back onto the AGB, giving rise to the
designation ``born-again'' AGB star \citep{iben:83a}. If this scenario is true, then the
intershell abundances in the models has to be brought into agreement with
observations. By introducing a more effective overshoot prescription for the
He-shell flash convection during thermal pulses on the AGB, dredge-up of carbon
and oxygen into the intershell can achieve this agreement \citep{herwig:99c}. Another strong
support for the born-again scenario was the detection of neon lines in optical spectra of
some PG1159 stars \citep{werner:94}. The abundance analysis revealed Ne=0.02, which is in good
agreement with the Ne intershell abundance in the improved stellar models.

If we do accept the hypothesis that PG1159 stars display former intershell
matter on their surface, then we can in turn use these stars as a tool to investigate
intershell abundances of other elements. Therefore these stars offer the unique
possibility to directly see the outcome of nuclear reactions and mixing
processes in the intershell of AGB stars. Usually the intershell is kept hidden
below a thick H-rich stellar mantle and the only chance to obtain information about
intershell processes is the occurrence of the third dredge-up. This indirect
view onto intershell abundances makes the interpretation of the nuclear and
mixing processes very difficult, because the abundances of the dredged-up
elements may have been changed by
additional burning and mixing processes in the H-envelope (e.g., hot-bottom
burning). In addition, stars with an initial mass below 1.5~\msun\ do not
experience a third dredge-up at all.

For completeness we note that the central stars of planetary nebulae of spectral
type [WC] are believed to be immediate progenitors of PG1159 stars,
representing the evolutionary phase between the early post-AGB and PG1159
stages. This is based on spectral analyses of [WC] stars which yield very
similar abundance results \citep{hamann:97}. We do not discuss the [WC] stars here
because the analyses of trace elements are much more difficult or even impossible
due to strong line broadening in their rapidly expanding atmospheres.

\begin{figure*}
\vspace{-2.8cm}
\begin{center}
\includegraphics[scale=.60]{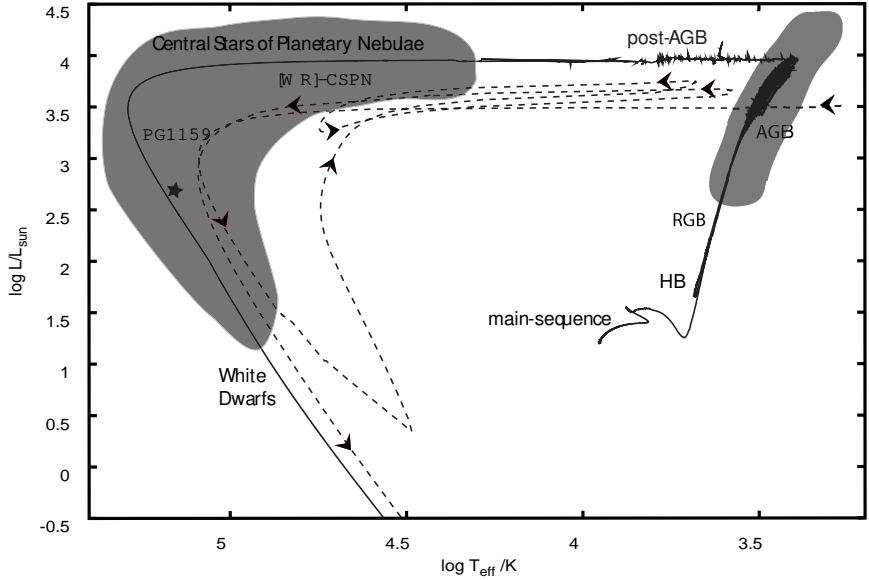}
\end{center}
\vspace{-6.3cm}
\caption{\label{fig:hrd} Complete stellar evolution track with an initial mass
of $2\msun$ from the main sequence through the RGB phase, the
HB to the AGB phase, and finally
through the post-AGB phase that includes the central stars of planetary nebulae
to the final WD stage. The solid line represents the evolution of a
H-normal post-AGB star. The dashed line shows a born-again evolution of the same
mass, triggered
by a very late thermal pulse, however, shifted by
approximately $\Delta \log \Teff = - 0.2$ and $\Delta \log L/\lsun = - 0.5$ for
clarity. The star shows the position of PG1159-035. }
\end{figure*}                                                                                        

\section{Three different late He-shell flash scenarios}

The course of events after the final He-shell flash is qualitatively different
depending on the moment when the flash starts. We speak about a very late
thermal pulse (VLTP) when it occurs in a WD, i.e. the star has turned
around the ``knee'' in the HRD and H-shell burning has already stopped (Fig.\,\ref{fig:hrd}). The
star expands and develops a H-envelope convection zone that eventually reaches deep
enough that H-burning sets in (a so-called hydrogen-ingestion flash). Hence H is destroyed and whatever H abundance
remains, it will probably be shed off from the star during the ``born-again''
AGB phase. A late thermal pulse (LTP) denotes the occurrence of the final flash
in a post-AGB star that is still burning hydrogen, i.e., it is on the horizontal
part of the post-AGB track, before the ``knee''. In contrast to the VLTP case, the bottom of the developing
H-envelope convection zone does not reach deep enough layers to burn H. The
H-envelope (having a mass of about $10^{-4}$~\msun) is mixed with a few times
$10^{-3}$~\msun\ intershell material, leading to a dilution of H down to about
H=0.02, which is below the spectroscopic detection limit. If the final flash
occurs immediately before the star departs from the AGB, then we talk about an
AFTP (AGB final thermal pulse). In contrast to an ordinary AGB thermal pulse the
H-envelope mass is particularly small. Like in the LTP case, H is just diluted
with intershell material and not burned. The remaining H abundance is relatively
high, well above the detection limit (H $\gappr$ 0.1).

There are three objects, from which we believe to have witnessed a (very) late
thermal pulse during the last $\approx$\,100 years. FG~Sge suffered a late flash
in 1894 \citep{gonzalez:98}. The star became rich in C and rare earth elements. It most probably was
hit by an LTP, not a VLTP, because it turned
H-deficient only recently (if at all, this is still under debate). As of today,
FG~Sge is located on or close to
the AGB.

V605~Aql has experienced a VLTP in 1917 \citep{clayton:97}. Since
then, it has quickly evolved back towards the AGB, began to reheat and is now in
its second post-AGB phase. It has now an effective temperature of the order
100\,000~K and is H-deficient.

Sakurai's famous object (V4334~Sgr) also experienced a VLTP, starting around
1993 \citep{duerbeck:96}. It quickly evolved back to the AGB and became H-deficient. Recent
observations indicate that the reheating of the star already began, i.e., its
second departure from the AGB might just have begun.

The spectroscopic study of FG~Sge and Sakurai's object is particularly
interesting, because we can observe how the surface abundances change with
time. The stars are still cool, so that isotopic ratios can be studied from
molecule lines \citep{pavlenko:2004} and
abundances of many metals can be determined. The situation is less favorable with the
hot PG1159 stars: All elements are highly ionised and for many of them no atomic
data are available for quantitative analyses. On the other hand, in the cool
born-again stars the He-intershell
material is once again partially concealed.

\section{Comparison of observed and predicted element abundances}

Abundance analyses of PG1159 stars are performed by detailed fits to spectral
line profiles. Because of the high \Teff\ all species are highly ionized and,
hence, most metals are only accessible by UV spectroscopy. Optical spectra
always exhibit lines from \ion{He}{ii} and \ion{C}{iv}. Only the hottest
PG1159 stars display additional lines of N, O, and Ne (\ion{N}{v}, \ion{O}{vi},
\ion{Ne}{vii}). For all other species we have utilized high-resolution UV spectra
that were taken with the Hubble Space Telescope (HST) and the Far Ultraviolet
Spectroscopic Explorer (FUSE). FUSE allows observations in the Lyman-UV
range ($\approx$\,900--1200~\AA) that is not accessible with HST, and this turned out to
be essential for many results reported here. 

\noindent
\emph{Hydrogen --} Four PG1159 stars show residual H with an abundance of
0.17. These objects are the outcome of an AFTP. All other PG1159 stars have
H $\lappr$ 0.1 and, hence, should be LTP or VLTP objects. 

\noindent
\emph{Helium, carbon, oxygen --} These are the main constituents of PG1159
atmospheres. A large variety of relative He/C/O abundances is observed. The
approximate abundance ranges are: He=0.30--0.85, C=0.15--0.60, O=0.02--0.20. The
spread of abundances might be explained by different numbers of thermal pulses
during the AGB phase.

\noindent
\emph{Nitrogen --} N is a key element that allows to decide if the star is the
product of a VLTP or a LTP. Models predict that N is diluted during an LTP so
that in the end N=0.1\%. This low N abundance is undetectable in the optical and
only detectable in extremely good UV spectra. In contrast, a VLTP produces
nitrogen (because of H-ingestion and burning) to an amount of 1\% to maybe a few percent. N
abundances of the order 1\% are found in some PG1159 stars, while in others it
is definitely much lower. 

\noindent
\emph{Neon --} Ne is produced from $^{14}$N that was produced by CNO burning. In the
He-burning region, two $\alpha$-captures transform $^{14}$N to $^{22}$Ne. Stellar evolution
models predict Ne=0.02 in the intershell. A small spread is
expected as a consequence of different initial stellar masses. Ne=0.02 was found in
early optical analyses of a few stars and, later, in a much larger sample
observed with FUSE \citep{werner:04}.

\noindent
\emph{Fluorine --} F was for the first time discovered by \citet{werner:05} in hot post-AGB stars; in
PG1159 stars as well as hydrogen-normal central stars. A strong absorption line
in FUSE spectra located at 1139.5~\AA\ remained unidentified until we found that
it stems from \ion{F}{vi}. The abundances derived for PG1159
stars show a large spread, ranging from solar to up to 250 times solar. This was
surprising at the outset because $^{19}$F, the only stable F isotope, is very fragile
and easily destroyed by H and He. A comparison with
AGB star models of \citet{lugaro:04a}, however, shows that such high F abundances in
the intershell can indeed be accumulated by the reaction
$^{14}$N($\alpha$,$\gamma$)$^{18}$F($\beta^+$)$^{18}$O(p,$\alpha$)$^{15}$N($\alpha$,$\gamma$)$^{19}$F,
the amount depends on the
stellar mass. We find a good agreement between observation and theory. Our
results also suggest, however, that the F overabundances found in AGB stars
\citep{jorissen:92} can only be understood if the dredge-up of F in the AGB
stars is much more efficient than hitherto thought.

\noindent
\emph{Silicon --} The Si abundance in evolution models remains almost
unchanged. This is in agreement with the PG1159 stars for which we could
determine the Si abundance.

\noindent
\emph{Phosphorus --} Systematic predictions from evolutionary model grids are not
available; however, the few computed models show P overabundances in the range 4--25
times solar (Lugaro priv. comm.). This is at odds with our spectroscopic measurements for
two PG1159 stars, that reveal a solar P abundance.

\noindent
\emph{Sulfur --} Again, model predictions are uncertain at the moment. Current
models show a slight (0.6 solar) underabundance. In strong contrast, we find a
large spread of S abundances in PG1159 stars, ranging from solar down to 0.01 solar.

\noindent\emph{Lithium --} Unfortunately, PG1159 stars are too hot to exhibit Li
lines because Li is completely ionised. If Li were detected then it must have been
produced during a VLTP. The discovery of Li in Sakurai's star is a strong
additional hint that it underwent a VLTP and not an LTP.

\noindent
\emph{Iron and Nickel --} \ion{Fe}{vii} lines are expected to be the strongest iron
features in PG1159 stars. They are located in the UV range. One of the most
surprising results is the non-detection of these lines in three examined PG1159 stars
(K1-16, NGC\,7094, PG1159-035). The derived upper abundance limits \citep[e.g.][]{werner:03} indicate that
iron is depleted by about 0.7--2 dex, depending on the particular object. Iron
depletions were also found for the PG1159-[WC] transition object Abell~78 as
well as for several PG1159 progenitors, the [WC] stars. Such high Fe depletions
are not in agreement with current AGB models. Destruction of $^{56}$Fe by
neutron captures is taking place in the AGB star intershell as a starting point
of the s-process; however, the resulting depletion of Fe in the intershell is
predicted to be small (about
0.2~dex). It could be that additional Fe depletion can occur during the late
thermal pulse. In any case, we would expect a simultaneous enrichment of nickel,
but up to now we were unable to detect Ni in PG1159 stars at all. While the
solar Fe/Ni ratio is about 20, we would expect a ratio close to the s-process
quasi steady-state ratio of about 3. Fittingly, this low ratio has been found in
Sakurai's (cool) LTP object.

\noindent
\emph{Trans-iron elements --}
The discovery of s-process elements in PG1159 stars would be
highly desirable. However, this is at present impossible due to the lack of atomic
data. From the ionization potentials we expect that these elements are highly
ionised like iron, i.e., the dominant ionization stages are \ion{}{vi-ix}. To
our best knowledge, there are no laboratory measurements of so highly ionised
s-process elements that would allow us to search for atomic lines in the
observed spectra. Such measurements would be crucial to continue the element
abundance determination beyond the current state.

\section{Conclusions}

It has been realized that PG1159 stars exhibit intershell matter on their
surface, which has probably been laid bare by a late final thermal pulse. This
provides the unique opportunity to study directly the result of nucleosynthesis
and mixing processes in AGB stars. Spectroscopic abundance
determinations of PG1159 photospheres are in agreement with intershell
abundances predicted by AGB star models for many elements (He, C, N, O, Ne, F,
Si). For other elements, however, disagreement is found (Fe, P, S) that points at possible
weaknesses in the evolutionary models.

\begin{acknowledgements}
ER and TR are funded by DFG (grant We 1312/30-1) and DLR (grant 50\,OR\,0201),
respectively. This work was funded in part under the auspices of the U.S.\
Dept.\ of Energy under the ASC program and the LDRD program (20060357ER) at Los
Alamos National Laboratory (FH). JWK is supported by the FUSE project, funded by
NASA contract NAS5-32985.  Figs.\,\ref{fighrd} and \ref{fig:hrd}
originally appeared in the Publications of the Astronomical Society of the
Pacific (Werner \& Herwig 2006).  Copyright 2006, Astronomical Society of the
Pacific; reproduced with permission of the Editors.
\end{acknowledgements}

\bibliographystyle{aa}

\end{document}